\documentclass[dvips,twocolumn]{article}\usepackage{graphicx}
\begin{document}

\title{On Simple Mean-Field Stochastic Model of Market Dynamics}
\author {Guennadi Saiko\footnote{Author email: gsayko@yahoo.com} \\  \small Dept. of Radiophysics and Solid-State Electronics, MIPT, 141700 Dolgoprudniy \textit{(Russia)}}
\date{\small June 30, 2003}
\maketitle

\begin{abstract}
We propose a simple stochastic model of market behavior. Dividing market participants into two groups: trend-followers and fundamentalists, we derive the general form of a stochastic equation of market dynamics. The model has two characteristic time scales: the time of changes of market environment ($t_m$) and the characteristic time of news flow ($t_n$). Price behavior in the most general case is driven by three stochastic processes, attributed to trend-followers, fundamentalists, and news flow, respectively. The model demonstrates the wide range of peculiarities which are typical in real markets: multiscale behavior, clustered volatility, weak correlations between the price changes on successive trading days, etc.
\end{abstract}

\section{Introduction}
Volatility and crashes of stock markets in recent years have become a source of inspiration for numerous researchers (see, e.g. [1,2,4] and references therein). Although the number of publications on this topic at least doubles each year, we probably can see only the tip of the iceberg. The major part of research takes place in secrecy - in private funds, banks, and investment companies; their results are strictly confidential and probably will never be published. Nevertheless, available data in open sources are also plentiful in empirical approaches [1,2], simulations of markets based on economical models [3,4], and analytical models [5-8]. 
 
The aim of this article is to construct a consistent approach, which can adequately describe market behavior. Because many models of the markets work well only until the market crashes, the necessary feature of a good model is to provide a possibility to explain these abnormal market phases. So, an adequate model must imitate normal market behavior as well as market peculiarities - turmoil.

The key point in understanding market behavior is probably that a log-periodicity, fat tails, cluster volatility, and other peculiarities of markets are quite similar [2] to those in the theory of critical phenomena and self-organization. Consequently, the basic feature of a market is that it is primarily a multi-body system. The complexity of investors and intricacy of their strategies are likely secondary factors - they do not affect the market as a whole. So, to describe market behavior we should study the multi-agent system rather than examine the decision making of an individual investor. 

The idea is to start from a rather general analytical model and then add complexity and make assumptions gradually. Because we consider that the applicability of mean field approach has not been drained yet, therefore we use this method.

\section{Model}
In the free market the price of an asset $p(t)$ is determined by the balance between supply and demand. Specifically, the return ($dp/p$) is proportional to a net order size - the difference between the number of demand orders ($d$) and supply orders ($s$) (see, for example [2].) Or, 
\begin{equation}
\frac{dp}{p}=K(d-s)dt
\end{equation}
here $K$ is the elasticity of the market, which is inversely proportional to the market size (the larger the market, the smaller is the impact of the particular imbalance.) 

We consider the market consisting of N participants or agents. The state of the agent $i$ is represented by  $\phi_i=\{-1,0,1\}$. Here $\phi_i=0$ corresponds to an inactive state (the agent does not participate in the market or waits), "-1" and "1" correspond to selling and buying states, respectively. 

To calculate the net order size $\Phi$ we must sum $\phi_i$ over all of the agents. Even though we do not know anything about the decision-making process of each agent, their huge number simplifies this problem - we can use the Central Limit Theorem. Thus, we can expect that the net order sizeat a given time $\Phi$ is a stochastic value that is distributed normally around its mean (unfortunately, this mean as well as the deviation can change (drift) during  time.) At this stage, the equation of market dynamics has a traditional form of a geometric Brownian motion, which was analyzed by Black and Scholes [9]:
\begin{equation}
\frac{dp}{p}=K(\mu dt+\sigma dW)
\end{equation}
where $K\mu$ is the expected return per unit time, $\sigma^{2}$ is the variance per unit time (volatility), and W is a Wiener process (if we assume independent increments).
However, this model does not describe all the peculiarities of the markets. To obtain a more realistic model we must make some additional assumptions about the market. In order to keep the model in a rather general form we make only one reasonable assumption (see, for example [10]), stating that the decision about participation in the market can be induced by market data (trend-followers or chartists) or external news (fundamentalists). 
This assumption allows us to decompose the expression for   $\Sigma\phi_i$ into two independent parts ($\Phi_{TF}$ and $\Phi_{F}$):

\begin{equation}
\Sigma\phi_i=\Phi_{TF} + S*News
\end{equation}  
where $\Phi_{TF}$ - some stochastic function which depends on price history, $News$ - is the flow of news related to this asset, $S$ - is the sensitivity of agents to the news.

Because both parts of Eq.3 are the sum of a large number of values $\{-1, 1\}$, we can make some predictions about $\Phi_{TF}$ and $S$ - they are classical random walk processes. Thus, both terms on the right side of Eq.3 are stochastic values normally distributed with some means and variances; we can model $\Phi_{TF}$ and $S$ by stochastic processes. It reflects the fact that the market environment changes permanently - agents with different strategies go in and out of the market. The only question is about the pace of these changes and what the characteristic time of these stochastic processes is (this is a question we will consider a little later.)

As a result, we have the equation for market dynamics:
\begin{equation}
\frac{dp}{p}=K(\mu dt + d\xi_{TF} +(\delta dt + d\xi_F)d\xi_N) 
\end{equation}
where $\mu$ and $\delta$ are market "bias"; $\xi_{TF}$ , $\xi_F$ , and $\xi_N$ are stochastic processes which describe trend-followers, fundamentalists, and news, respectively. We introduce here $\mu$ and $\delta$ in order to shift the stochastic processes to Gaussian ones with means which are equal to zero.

\section{Results and Discussion}
The derived equation describes the market dynamics in a quite general form. However, we must pay for this generality - to mimic market behavior in this approach we use three  different stochastic processes instead of one in traditional models.

Fortunately, the situation is not too pessimistic and to explore the model we can simplify this equation in some important cases. As we mentioned, the first term (conditioned by trend-followers) and the sensitivity to the news (conditioned by fundamentalists) are stochastic processes. However, the typical scale of these processes is conditioned by large changes in market participation. So, this time scale $(t_m)$ differs from the time scale of news $(t_n)$. Using this observation, we can simplify the equation in two major situations: $t_m \ll t_n$ (very slow news and rapid investors) and $t_m \gg t_n$ (very rapid news and slow investors).

After averaging the right part of Eq.4, we can explore the market in two cases: $t > t_n$ and $t > t_m$. These results are depicted in Tab.1 and Tab.2. Here we use the following notation: $A$, $B$, and $C$ are some market constants at a particular time. We also substituted the stochastic processes $\xi_{TF}$, $\xi_F$, and $\xi_N$ with $\sigma_{TF}\xi_1$ , $\sigma_F\xi_2$, and $\sigma_N\xi_3$, respectively, where $\xi_1$, $\xi_2$, and $\xi_3$ are stochastic processes with the standard deviation equals to 1.

\begin{table}[ht]
\begin{tabular}{c|c}
\multicolumn{2}{c}{Tab 1: Asymptotic behavior of $log(p)$ if $t_m \ll t_n$}\\
 \hline
$t > t_n$ & $K(\mu t + \delta \sigma_N \xi_3)$ \\
\hline
$t > t_m$ & $K(\mu t + \sigma_{TF}\xi_1 +(\delta t + \sigma_F\xi_2)\sigma_N C)$ \\
\hline
\end{tabular}
\end{table}
\begin{table}[ht]
\begin{tabular}{c|c}
\multicolumn{2}{c}{Tab 2: Asymptotic behavior of $log(p)$ if $t_m \gg t_n$}\\
 \hline
$t > t_n$  & $K(A t + B \sigma_N \xi_3)$\\
\hline
$t > t_m$  & $K(\mu t +  \sigma_{TF} \xi_1)$\\
\hline
\end{tabular}
\end{table}

It is easy to see that in the case $t_m \gg t_n$ (which probably is typical for real markets) the short scale and large scale dynamics are determined by the different stochastic processes. The market is news-driven on the short scale and trend follower-driven on the large scale.

In the opposite case ($t_m \ll t_n$), the market is news-driven on the short scale and participant-driven on the large scale.

Another interesting situation is the market dominance of trend-followers or fundamentalists. In the former case, market dynamics is conditioned primarily by the first two terms in Eq.4. As a result, we have the directed dynamics (if $\mu\neq 0$) with a time scale conditioned by market participation ($t_m$).

In the fundamentalists' market case, dynamics is determined by two concurrent stochastic processes - market participation ($t_m$) and news flow ($t_n$). It results primarily in a side trend (if the news flow is "unbiased").

To investigate the statistical properties of the model we should specify the stochastic processes. Not taking into account the turbulent situations, where the high correlations between market agents are possible (herd behavior, [7]), we can expect that market participants' changes and news have independent increments. So, we can imitate all three stochastic processes by Wiener ones.

Because the model has two different time scales ($t_m$ and $t_n$), it is important to explore it with different $t_m$ and $t_n$. In Fig.1 a-d we can see the typical evolution of the system with different characteristic times $t_n$ and $t_m$. It is easy to see the variety of patterns generated by this model: clustered volatility, multiscaling, etc.
\begin{figure}[ht]
	\begin{center}	
\includegraphics[width=3.2in]{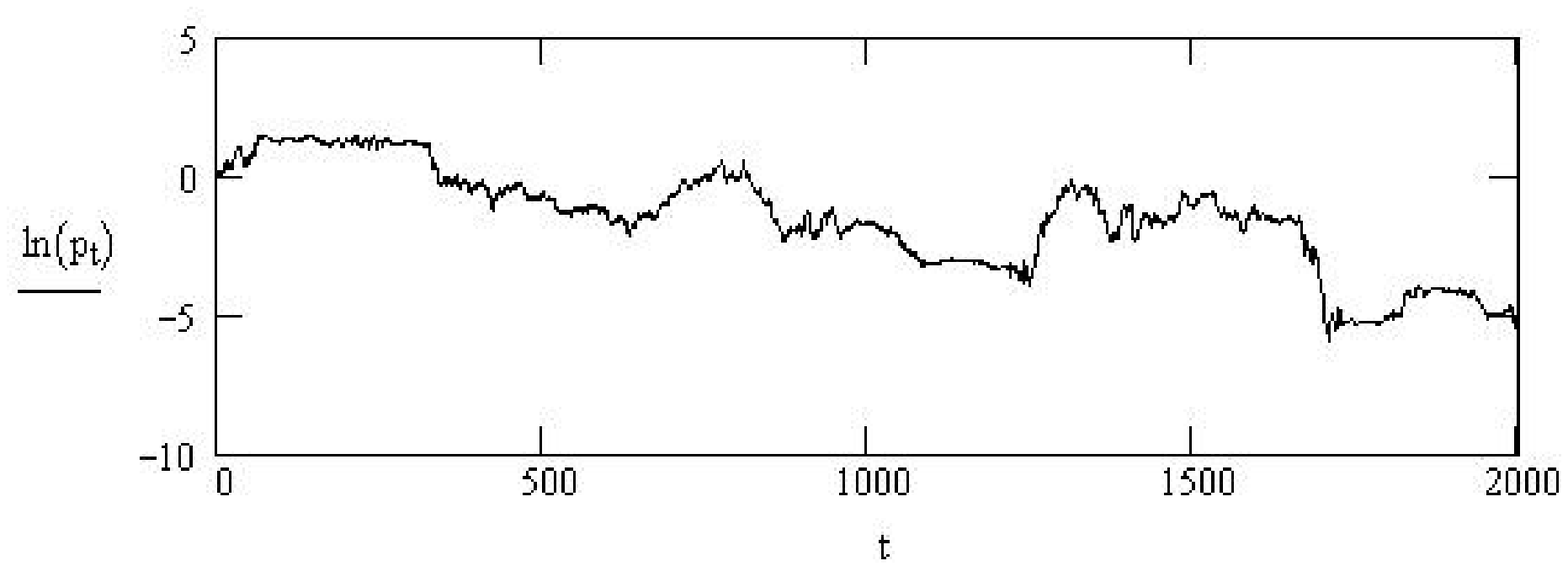}
\includegraphics[width=3.2in]{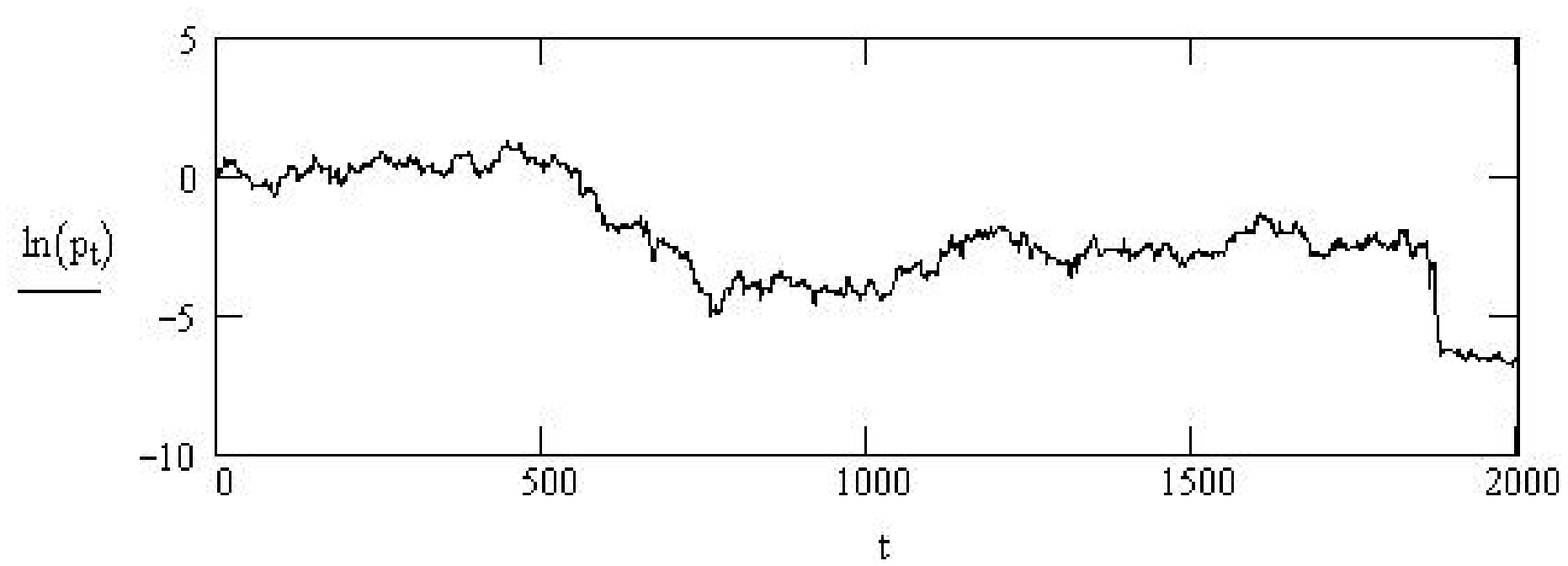}
\includegraphics[width=3.2in]{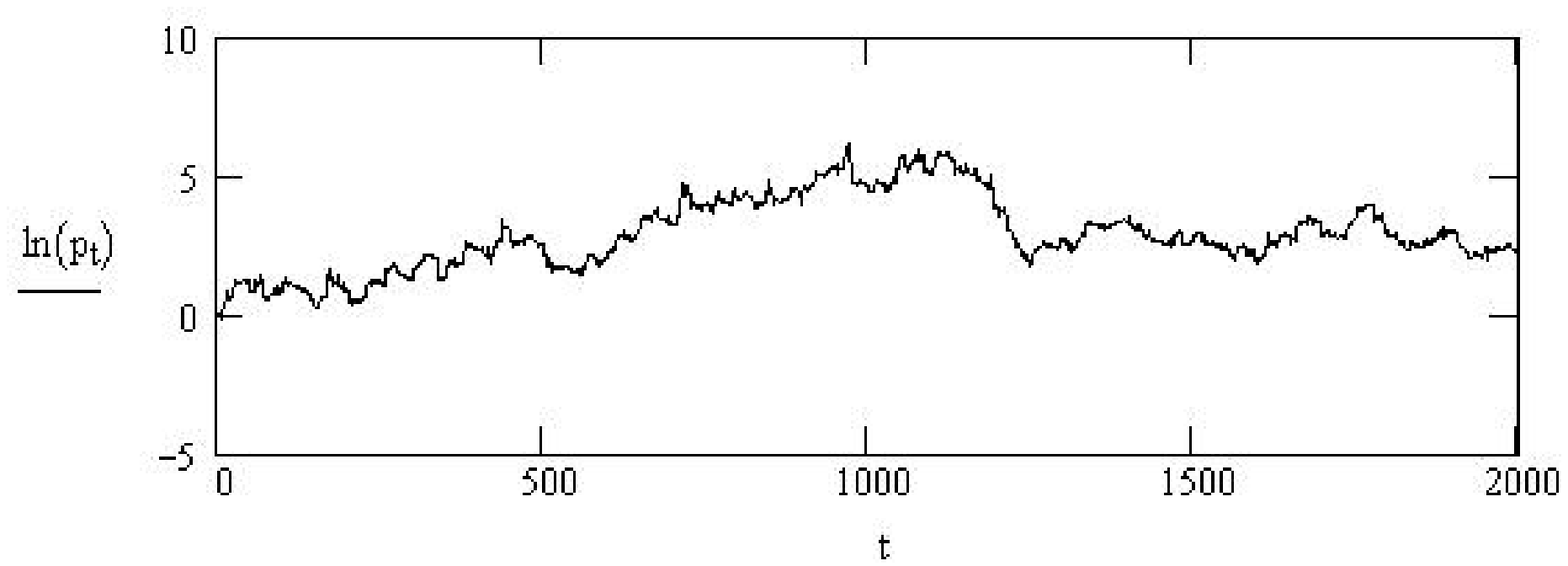}
\includegraphics[width=3.2in]{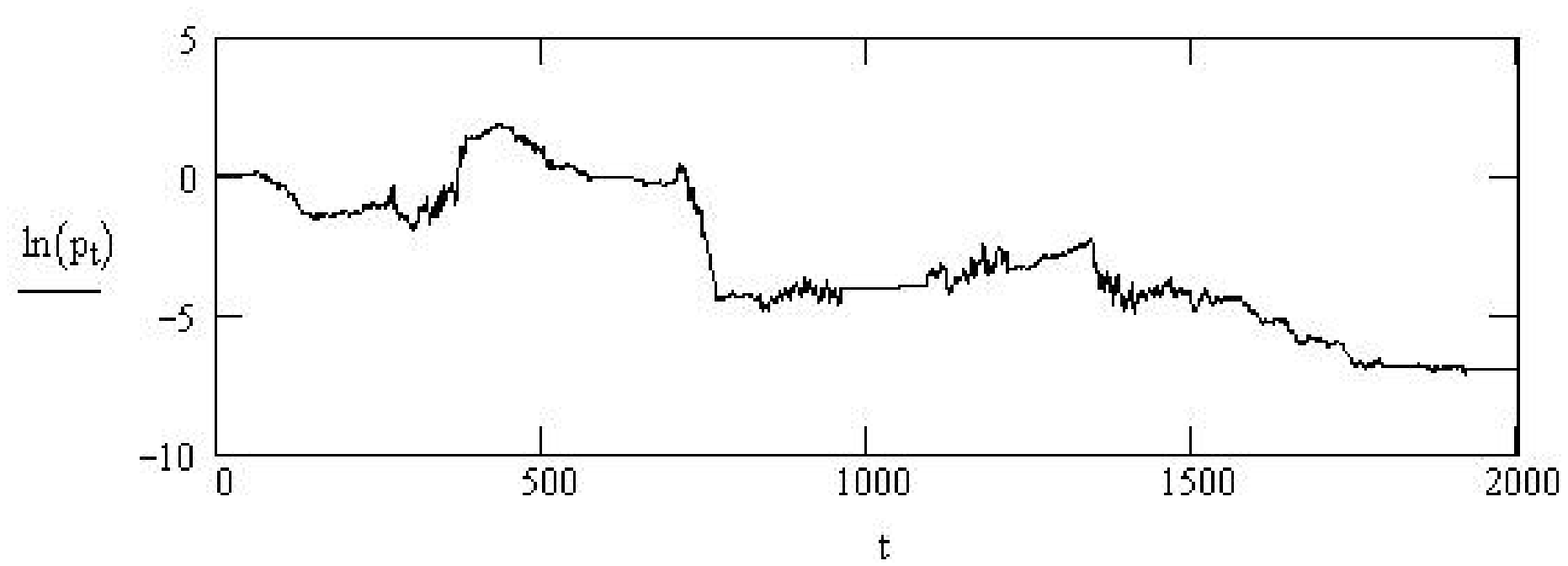}
	\end{center}
	\caption{Time series of the typical evolution of the market price $p$ at various characteristic times $t_m$ and $t_n$. From the top to the bottom: a) $t_m=64$, $t_n=1$, b) $t_m=4$, $t_n=1$, c) $t_m=1$, $t_n=4$, d) $t_m=1$, $t_n=64$.}
\end{figure}

To test the data generated by the model, we use standard methods: skewness, kurtosis, Hurst exponent, analysis of drawdown. The results show that the model  generates outliers, fat tails, and other peculiarities of real markets. 

In particular, Fig.2 depicts the dependence of the skewness and the kurtosis of returns distribution vs. the variance of "participant noise". While the kurtosis is close to zero if the noise is negligible, the high values of kurtosis (and outliers, respectively) can be observed with the increase of the noise.

\begin{figure}[ht]
\centering
\includegraphics[width=1.5in]{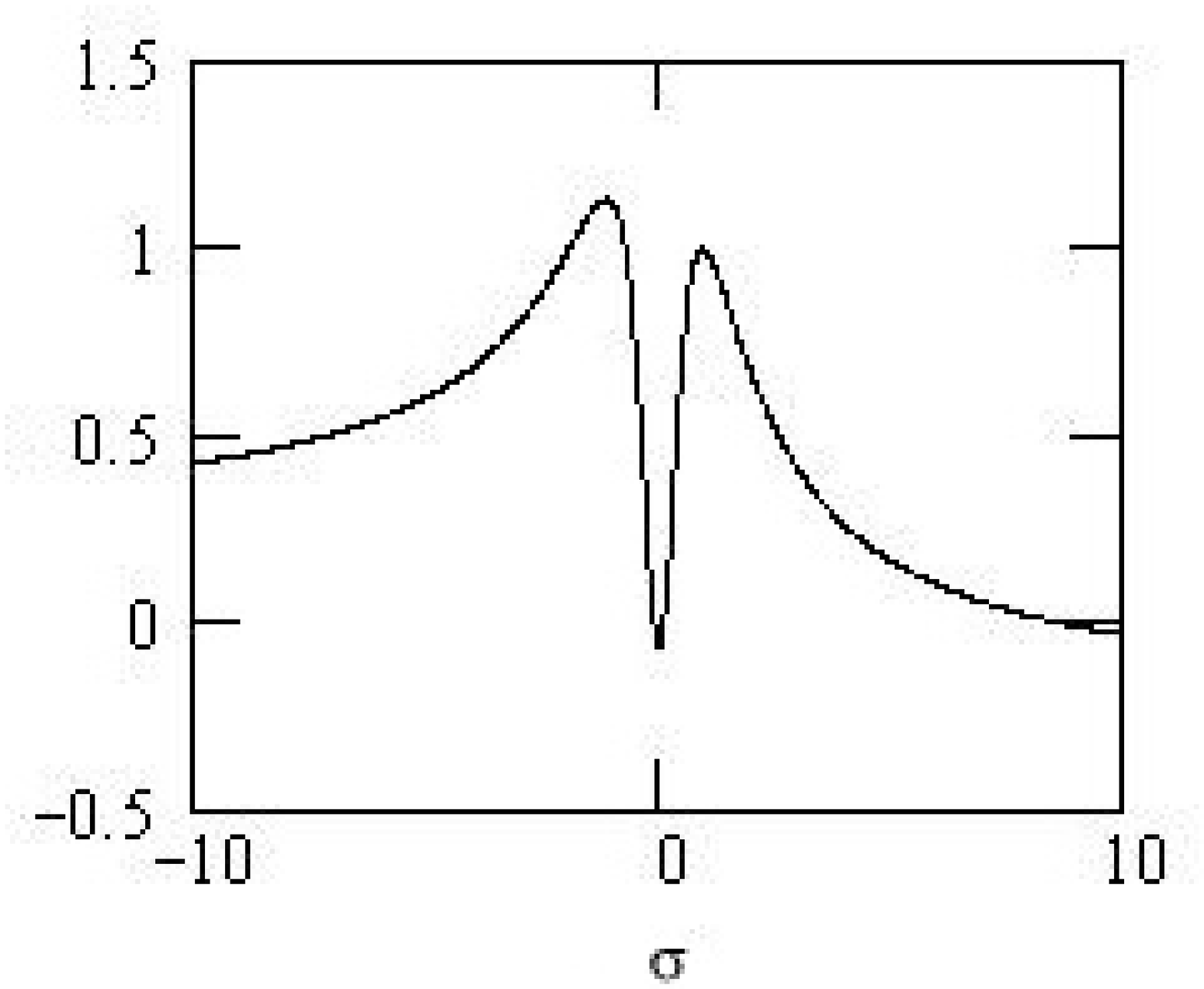}
 \includegraphics[width=1.5in]{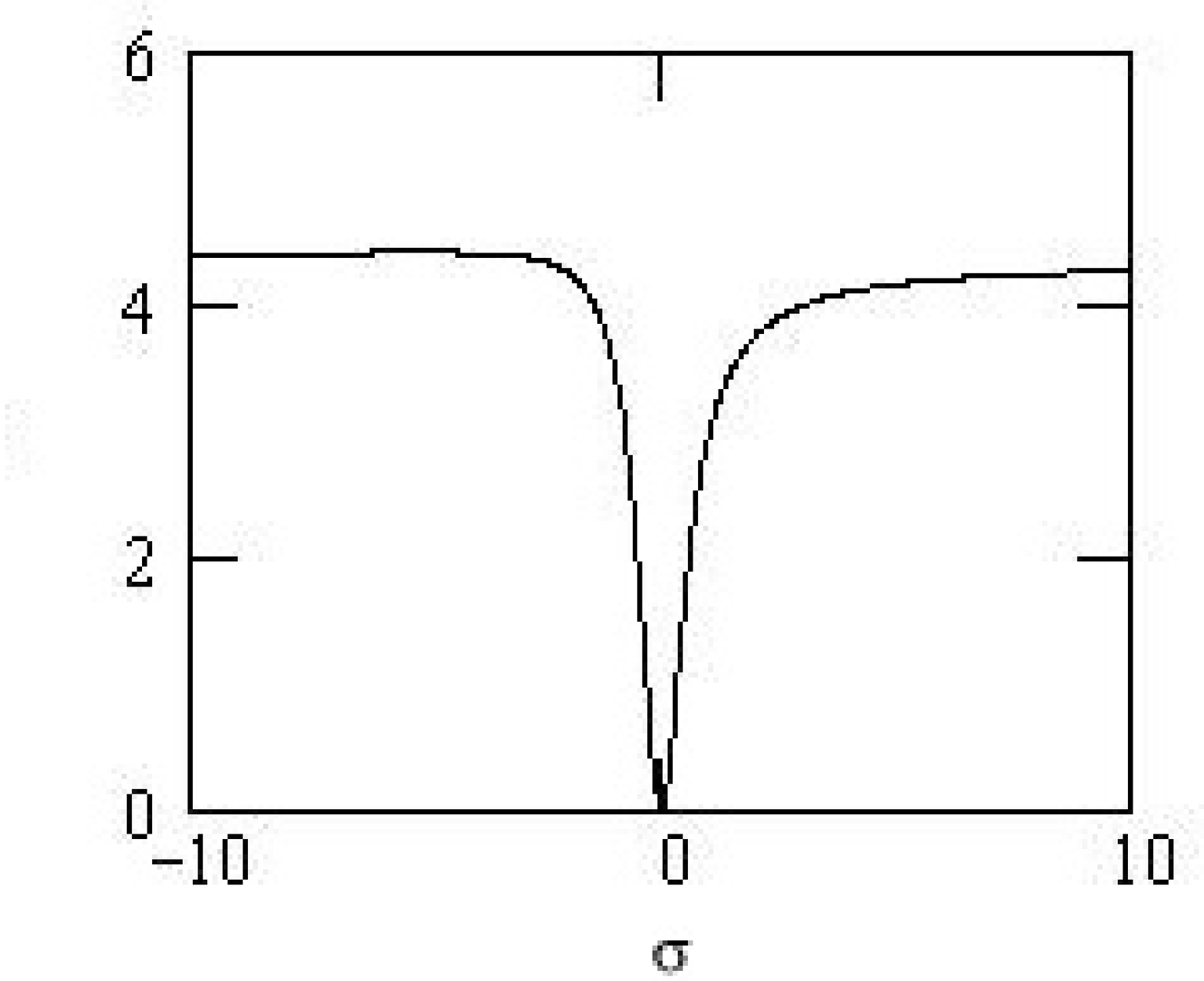}
 \vspace{0.2in}
\caption{The typical dependance of the skewness and kurtosis of the return distribution on the market participation parameter $\sigma$ ($\sigma=\sigma_{F}=\sigma_{TF}$)}
\end{figure}

\section{Conclusions}
Thus, we can conclude that the model developed in this article exhibits the following features:

\textbf{Different time scales} - the model like real markets behaves differently on the short and large time scales. Consequently, different stochastic processes can lead to fat tails on the short scales (intraday trades) and crossover to Gaussian-distributed returns on larger scales (weeks). The presence of a few different time scales can also explain other interesting phenomena of the real markets: clustered 
volatility and multiscale behavior.

\textbf{Return distributions demonstrate leptocurtic behavior} - all analyzed time series with the market participation parameter $\sigma\neq0$ have kurtosis bigger than 0 - they demonstrate typical for real markets "high peaks" and "fat tails".

\textbf{Trend-follower vs. fundamentalist behavior} - if a trend-followers' market is primarily a directed market, a fundamentalists' market demonstrates mainly the lateral trend.

\textbf{Stabilization of a market with the growth of a number of market participants} -  such as the elasticity of the market ($K$) is inversely proportional to the number of market participants ($1/N$) and the net order size is a random walk process (which lead to the $N^{1/2}$ behavior), the amplitude of returns (volatility) is proportional to $N^{-1/2}$. It reflects the obvious property of markets - the increase of the number of participants stabilizes and smoothes the market (if other factors are constant). 

In summary, we have derived the stochastic equation of market dynamics. A brief look at its analytical sequences and computational results shows the variety of possible patterns, crossovers, and characteristic features which are typical for real markets. The equation has a rather general form and for its practical use one should make some additional assumptions. However, it is beyond the scope of this work and will be considered in following articles.

\end{document}